\documentclass{article}
\usepackage{amssymb}

\begin{document}

\title{Mathematics of the NFAT signalling pathway} 

\author{Alan D. Rendall\\
Max-Planck-Institut f\"ur Gravitationsphysik\\
Albert-Einstein-Institut\\
Am M\"uhlenberg 1\\
14476 Potsdam, Germany}

\date{}

\maketitle

\begin{abstract}
This paper is a mathematical study of some aspects of the signalling pathway 
leading to the activation of the transcription factor NFAT (nuclear factor
of activated T cells). Activation takes place by dephosphorylation at multiple
sites. This has been modelled by Salazar and H\"ofer using a large system
of ordinary differential equations depending on many parameters. With the help 
of chemical reaction network theory we show that for any choice of the 
parameters this system has a unique stationary solution for each value of
the conserved quantity given by the total amount of NFAT and that all 
solutions converge to this stationary solution at late times. The 
dephosphorylation is carried out by calcineurin, which in turn is activated 
by a rise in calcium concentration. We study the way in which the dynamics
of the calcium concentration influences NFAT activation, an issue also 
considered by Salazar and H\"ofer with the help of a model arising from work 
of Somogyi and Stucki. Criteria are obtained for convergence to equilibrium of 
solutions of the model for the calcium concentration. 
\end{abstract}

\section{Introduction}

The phenomena modelled mathematically in this paper are mechanisms 
which are part of the way the immune system works at the molecular level.
For background on immunology the reader is referred to \cite{murphy} or 
\cite{roitt}.
T cells are among the most important components of the immune system. They have 
the task of recognizing certain antigens and reacting appropriately. More
precisely, a T cell recognizes a peptide (small protein) in combination with
an MHC (major histocompatibility complex) molecule. The recognition takes 
place through a surface molecule, the T cell receptor. In what follows 
attention will be confined to T helper (Th) cells although some of the 
statements made may also apply to other types of T cells. In order for the T 
cell to be activated a second signal is also necessary. This comes from
another surface molecule, CD28, which recognizes the molecules B7.1 and B7.2 
on the antigen presenting cell carrying the peptide-MHC complex. The
information about these recognition events is propagated to the nucleus 
through various signalling pathways. The result is that the transcription
factors NFAT, NF$\kappa$B and AP-1 bind to the DNA, leading to the 
production of the cytokine IL-2 (interleukin 2). Much remains to be learned 
about these signalling pathways and mathematical modelling has a great 
potential to contribute to obtaining a better understanding of them.

In the following attention will be concentrated on the part of the 
signalling network relating to NFAT (nuclear factor of activated T cells).
It should be noted that although the abbreviation NFAT refers to
T cells this transcription factor is important for signalling in many other
types of cells. There are five different NFAT molecules and the one of 
relevance in what follows is that known as NFATc2 or NFAT1. 
A model for NFAT signalling in T cells was introduced by
Salazar and H\"ofer \cite{salazar03}. In fact they only deal with part of
the pathway. An important stage in signalling is when there is a flow of 
calcium ions into the cytosol. During the activation of T cells this occurs 
when IP${}_3$ (inositol 1,4,5-trisphosphate) binds to receptors in the 
endoplasmic reticulum (ER), opening calcium channels and thus allowing calcium 
ions to flow down their concentration gradient. This can be simulated 
experimentally by treating the cells with ionomycin, which leads to transport
of calcium ions across membranes. The first step in the NFAT pathway 
included in the work of \cite{salazar03} and the first one to play a role in 
what follows, is this increase in the calcium concentration. The calcium binds 
to calcineurin, partly activating it. It also binds to calmodulin, which can 
then complete the activation of calcineurin. The activated calcineurin removes 
phosphate groups from NFAT, which is present in phosphorylated form in the 
cytosol of resting cells. The NFAT then undergoes a conformational change and
moves to the nucleus where it can bind to DNA. The main model in  
\cite{salazar03} (which will be called the SH model in what follows) describes 
the dephosphorylation of NFAT and its transport between the cytosol and the 
nucleus. A subsidiary model describes the calcium influx. The aim of this 
paper is to obtain a deeper mathematical understanding of these models.

The SH model is a system of $4N+4$  equations and contains $10N+4$ 
parameters, where $N$ is the number of phosphorylation sites. The case of 
interest for NFAT is $N=13$ so that there are 56 equations and 134 parameters. 
Due to the large number of variables involved it might seem difficult to 
analyse the dynamical behaviour of general solutions of this system. Chemical 
reaction network theory (CRNT) \cite{feinberg80} is a general tool for 
attacking this type of problem and it turns out to be very effective in this 
case. As we will show, one of its strongest theorems, the Deficiency Zero 
Theorem, can be applied to this system. The result is that for a given total 
amount of NFAT there is a unique stationary solution of the system and that 
every other solution converges to the stationary solution.   

The SH model describes the dephosphorylation of NFAT when the concentration of 
activated calcineurin is constant. In fact the calcium influx which leads to 
the activation of NFAT is a dynamical process. To assess the applicability of
the SH model it is desirable to know whether the calcium concentration tends
to a constant value at late times. This process is modelled in \cite{salazar03}
by a two-dimensional dynamical system. It will be shown that for certain
subsets of the parameter space for this model the solutions do converge to
a stationary solution. It is also shown that when this happens the 
long-time behaviour of the amounts of the different forms of NFAT 
occurring in the SH model is that they converge to the values they converge to 
in the SH model with an appropriate choice of parameters. 

The paper is organized as follows. Section \ref{crnt} contains some basic 
material about chemical reaction network theory. The dynamical analysis of
the SH model for NFAT phosphorylation with constant stimulation is in 
section \ref{nfatdynamics}. The dynamics of the calcium influx is 
investigated in section \ref{calcium}. The last section of the main text
gives conclusions and an outlook. In an appendix the SH model is compared 
with a model for the NFAT signalling pathway defined in \cite{fisher06}.

\section{Chemical reaction network theory}\label{crnt}

Chemical reaction network theory is a collection of methods for studying the
dynamics of solutions of ordinary differential equations modelling systems 
of chemical reactions. Some concepts of this theory will now be reviewed. In 
CRNT the basic objects are finite sets $\cal S$ of species, $\cal C$ of 
complexes and $\cal R$ of reactions. The elements of $\cal C$ are formal 
linear combinations of elements of $\cal S$ with positive integer coefficients 
while the elements of $\cal R$ are ordered pairs of elements of $\cal C$.
The number of elements of $\cal S$, $\cal C$ and $\cal R$ are denoted by $m$,
$\bar n$ and $r$ respectively. The set $\cal S$ consists of the substances 
taking part in the chemical reactions and in the example of the SH system it 
consists of $4(N+1)$ states of NFAT. The complexes are the combinations of 
species occurring on the left and right hand sides of the reactions. In the 
SH model each complex is just a single species. The reactions are ordered 
pairs of complexes, each representing the input and output of one
reaction. In the case of the SH model they can be identified with ordered 
pairs of species. (Note that a reaction and the reverse reaction are counted 
separately.) Given a reaction network let $c_s$ be the concentration of the 
species with index $s$. Consider a system of ordinary differential equations 
of the form
\begin{equation}\label{generalkinetics}
\frac{dc_s}{dt}=f_s(y)=\sum_{(y,y')\in {\cal R}} r(y,y')(y_s'-y_s).
\end{equation}
Here $r(y,y')\ge 0$ are the reaction rates. They are assumed non-negative.
The choice of these functions is often referred to as the kinetics. In this 
section only the most standard choice of kinetics will be considered. This 
is mass-action kinetics, where
\begin{equation}
r(y,y')=k_{yy'}c^y.
\end{equation}
Here $k_{yy'}$ are positive constants called the rate constants and
$c^y=\prod_{s\in {\cal S}}c_s^{y_s}$. The equation (\ref{generalkinetics}) will be
abbreviated to $\dot c=f(c)$. Here $c$ is a vector of concentrations $c_s$ and 
so is a point of ${\bf R}^m$. The positive and non-negative orthants are 
defined to be the sets of points of ${\bf R}^m$ whose coordinates are positive 
and non-negative, respectively. The quantity $c$ is said to be positive 
(non-negative) if it lies in the positive (non-negative) orthant. Because of 
its interpretation in terms of concentrations $c$ should be non-negative
in order to be of relevance for applications.

The positive and non-negative orthants are invariant under the evolution
defined by the ordinary differential equations of a chemical reaction 
network. The invariance of the non-negative orthant follows from that of the 
positive orthant by continuity. The invariance of the positive orthant is a 
consequence of a lemma which will now be proved. (Cf. Lemma II.1 of 
\cite{sontag01} for a similar result.)

\vskip 10pt\noindent
{\bf Lemma 1} Consider a solution $c_s(t)$ of (\ref{generalkinetics}) with
the coefficients $r(y,y')$ being given by mass-action kinetics. If $c_s(t_0)>0$
for some $s$ and some time $t_0$ then $c_s(t)>0$ for all $t\ge t_0$ for 
which the solution exists.

\noindent
{\bf Proof} If the statement of the lemma is false then it can be assumed 
that $c_s(t_1)=0$ for some $t_1>t_0$. The time $t_1$ can be chosen so that 
$c_s(t)>0$ for all $t<t_1$. The quantity $c_s$ satisfies an equation of the 
form 
\begin{equation}
\frac{dc_s}{dt}=-f_-(c)c_s+f_+(c)
\end{equation}
where $f_+$ is non-negative. Since $c_s$ is positive on the interval 
$[t_0,t_1)$ the inequality 
\begin{equation}
\frac{d}{dt}(\log c_s)\ge -f_-(c)
\end{equation} 
holds. Integrating this equation and exponentiating gives
\begin{equation}
c_s(t_1)\ge c_s(t_0)\exp \left(-\int_{t_0}^{t_1}|f_-(c(t))|dt\right).
\end{equation} 
The integral in this expression is finite and so the inequality implies that
$c_s(t_1)>0$, a contradiction. This completes the proof of the lemma.

The reaction network can be represented as a directed graph where the 
vertices are the complexes and edges represent reactions. The reaction network 
is said to be weakly reversible if whenever it is possible to link the complex 
$y$ to the complex $y'$ by a sequence of reactions it is also possible to link 
$y'$ to $y$ in the same way. In the case of the SH system the network is weakly 
reversible since in fact every reaction is reversible. The connected 
components of the reaction graph are called linkage classes and their number 
is denoted by $l$. In the case of the SH model $l=1$. An important object is 
the stoichiometric matrix $\bar N$. Its columns correspond to the reactions 
belonging to the network. The entries in a column are defined by the sums of 
coefficients of the different species in the complexes occurring in the 
reaction, with the coefficients on the left hand side being counted negatively
and the coefficients on the right hand side positively. In other words, these 
are the net number of molecules of each species produced in the reaction. It 
is an $m\times r$ matrix. The cosets of the form $c+{\rm im}\ \bar N$ are 
called stoichiometric compatibility classes and are invariant under the flow 
of the system. The intersection of a stoichiometric compatibility class with 
the non-negative orthant is called a reaction simplex and is also invariant 
under the flow by Lemma 1. The rank of $\bar N$ (i.e. the dimension of the 
stoichiometric compatibility classes) is denoted by $s$. The 
deficiency of the network is defined by $\delta=\bar n-l-s$. 
The following is part of the Deficiency Zero Theorem which was first proved in
\cite{horn72a}, \cite{horn72b} and \cite{feinberg72b}.

\noindent
{\bf Theorem 1} Let $\dot c=f(c)$ be the system of ordinary differential 
equations defined by a chemical reaction network by means of mass-action 
kinetics. If the network is weakly reversible and of deficiency zero then 
there is a unique positive stationary solution in each stoichiometric 
compatibility class. The solution is asymptotically stable within its class. 

An important part of the proof of Theorem 1 is to show that there exists a
Lyapunov function $L(c)$ which is non-increasing along solutions and 
strictly decreasing along all positive solutions except for the stationary
solution. This means that the stationary solution is the only possible positive
$\omega$-limit point of a positive solution. The function $f$ can be written
in the form $Yg$ where $Y$ is called the complex matrix. Its columns are in
one to one correspondence with the complexes and the entries in the column 
corresponding to the complex $y$ are the components $y_s$. In the case of the 
SH system $Y$ is just the identity. If $c_*$ is a stationary solution then
$f(c_*)=0$. If in addition $g(c_*)=0$ the stationary solution is called complex
balanced. Evidently any stationary solution of the SH system is complex 
balanced. In fact it is a consequence of the proof of Theorem 1 that for
a system satisfying the assumptions of that theorem the intersection of the
kernel of $Y$ with the image of $\bar N$ is $\{0\}$, so that any
stationary solution of a system of that type is complex balanced.

\section{The model for NFAT phosphorylation}\label{nfatdynamics}

The basic variables in the model of \cite{salazar03} are amounts of
different forms of NFAT. Using amounts rather than concentrations 
avoids introducing extra factors of the ratio of the volumes of the two
compartments. This is just a matter of mathematical convenience.
Phosphate groups can be attached to this molecule at
up to $N=13$ sites. The index $n$ will be used to denote the number of 
phosphate groups and runs from zero to $N$. It is assumed that the phosphate 
groups are bound to the sites in a certain order and are removed in the 
reverse order. There are thus $N$ phosphorylation states in total. Each of 
these has an active and an inactive form. Each of them occurs in the cytosol
and in the nucleus. This gives a total of $4(N+1)$ variables decribing the 
amounts of the different substances. The 
processes of attaching a phosphate group and the conformational change between 
the active and inactive forms are reversible. It is necessary to prescribe 
$6N+2$ rate constants to describe the reactions in a given compartment. 
However the rate constants describing the transitions between active and 
inactive forms are chosen to be the same in both compartments. Rate constants 
are also required to describe the transport processes between the two 
compartments - the active form is transported into the nucleus and the 
inactive form out of the nucleus. It is assumed that there is just one 
rate constant for each of these two processes, independent of the 
phosphorylation state. A diagram of this reaction network can be found in 
\cite{salazar03}, Fig. 1. Assuming mass-action kinetics leads to a system of 
ordinary differential equations for the time evolution of the amounts
of the different substances.

The unknowns are as follows. The amount of active NFAT in the cytoplasm 
with $n$ phosphorylated residues is denoted by $a_n$, $n=0,1,\ldots ,N$. The 
amount of the corresponding inactive form is denoted by $i_n$. The 
amounts of these substances in the nucleus are denoted by $A_n$ and 
$I_n$. All these quantities are supposed non-negative.
Mass-action kinetics is assumed. For $1\le n\le N-1$ the dynamical 
equations for amounts in the nucleus are
\begin{eqnarray}
&&\frac{dA_n}{dt}=K_{n-1}A_{n-1}-K_nA_n+C_nA_{n+1}-C_{n-1}A_n\nonumber\\
&&+l_n^-I_n-l_n^+A_n+da_n,\\
&&\frac{dI_n}{dt}=K_{n-1}'I_{n-1}-K_n'I_n+C_n'I_{n+1}-C_{n-1}'I_n\nonumber\\
&&+l_n^+A_n-l_n^-I_n-fI_n
\end{eqnarray}
with rate constants $K_n$, $K_n'$, $C_n$, $C_n'$, $l_n^+$, $l_n^-$,  $d$ and 
$f$. Evolution equations for the cases $n=0$ and $n=N$ can be obtained by 
taking equations formally identical to those above with the conventions that
$K_{-1}=C_{-1}=K_{-1}'=C_{-1}'=0$ and that $C_N=C'_N=K_N=K_N'=0$. Analogous 
evolution equations for the quantities $a_n$ and $i_n$ can be obtained as 
follows. Replace $A_n$ and $I_n$ by $a_n$ and $i_n$ respectively everywhere 
except in the last term of each equation. Reverse the sign in the last term
of each equation. Replace $C_n$, $C_n'$, $K_n$, $K_n'$ by $c_n$, $c_n'$, $k_n$
and $k_n'$. It is stated in \cite{salazar03} that the 
transport processes are much slower than the reactions within each 
compartment. On a heuristic level this can be imported into the mathematics 
by assuming that the coefficients $d$ and $f$ are very small. It may be hoped 
that solutions of the full system can be approximated by solutions of the 
system obtained by setting $d=f=0$. In the latter system the equations 
describing amounts in the nucleus and the cytoplasm decouple. For this 
reason it is referred to in what follows as the decoupled system. To analyse 
this system it is enough to analyse the subsystems describing the dynamics in 
each compartment. 

The SH system satisfies $\bar n=4(N+1)$ and $l=1$. In order to show that 
Theorem 1 applies it therefore suffices to show that the rank of $\bar N$ is 
$4N+3$. This rank is equal to $4(N+1)$ minus the dimension of the kernel of
$(\bar N)^{T}$. The conditions for a vector to lie in this kernel are easy 
to analyse. Some of the conditions imply that all the components of the 
vector corresponding to amounts of substances in the cytosol are 
equal and that those corresponding to amounts of substances in the
nucleus are equal. Finally elements corresponding to the amounts of 
the same substance in both compartments are equal. Thus the kernel is 
one-dimensional and the rank of the stoichiometric matrix is $4N+3$. It 
follows that the deficiency is zero and Theorem 1 applies to the SH system.    

Theorem 1 leaves open the question whether under its hypotheses every
solution converges to a stationary solution. It will be shown below that this
is the case for the SH system. Note first that the total amount of
NFAT, which is the sum of all the variables in the SH system, is a conserved
quantity. Thus each reaction simplex is compact. It follows that all solutions 
exist globally in time towards the future and the $\omega$-limit set of any 
positive solution is connected. Combining this
with the remark about $\omega$-limit points made previously shows that if a
solution does not converge to the stationary solution its $\omega$-limit set
must be contained in the boundary of the positive orthant.

Suppose now that a positive solution $c$ has an $\omega$-limit point in the 
boundary of the positive orthant. There is a solution of the system, say 
$c_\infty$, passing through that $\omega$-limit point. The range of $c_\infty$
is contained entirely in the boundary. By Lemma 1 the number of non-zero 
components of $c_\infty$ can never decrease. It might a priori increase but
in any case it will be constant after a finite time. Thus when considering 
late-time behaviour it may be assumed without loss of generality than 
$(c_\infty)_s(t)$ is non-zero for $1\le s\le k$ and identically zero for 
$k+1\le s\le m$. Here $0\le k<m$. In fact $k>0$ since the sum of the variables 
$(c_\infty)_s$, which is the total amount of NFAT in the cell, is conserved. If
$k+1\le s\le m$ then $(\dot c_\infty)_s=0$. There are no negative contributions 
to $(\dot c_\infty)_s$ since reactions having species $s$ on their left hand 
side are not active. If there is a link from species $s$ to species $s'$ in 
the reaction network with non-zero concentration then a positive contribution 
to $(\dot c_\infty)_s$ results. It follows that if $s'$ is adjacent to $s$ in 
the network then $k+1\le s'\le m$. Since the reaction graph is connected this 
implies that all $c_s$ vanish identically, a contradiction. Thus it has been 
proved that there can be no $\omega$-limit points on the boundary and the 
following result is obtained:
 
\noindent
{\bf Theorem 2} Let $\dot c=f(c)$ be the system of Salazar and H\"ofer. There
is a unique stationary solution $c_*$ in each stoichiometric compatibility
class and each positive solution converges to a stationary solution as
$t\to\infty$.

\vskip 10pt
It has been conjectured that under the hypotheses of Theorem 1 every solution
converges to a stationary solution as $t\to\infty$. This is known as the
global attractor conjecture \cite{craciun09} and has recently been proved in 
the case that there is only one linkage class by Anderson \cite{anderson11}. 
Theorem 2 could be deduced from the result of \cite{anderson11} but it has 
been shown here that there is a much easier proof in this relatively simple 
case. 

There are many different ways in which multiple phosphorylation can be 
organized and this can give rise to many systems related to the SH model.
The phosphorylation is said to be processive if an enzyme which binds its
substrate once phosphorylates several sites before dissociating. It is said 
to be distributive if only one site per binding event is phosphorylated.
One type of distributive phosphorylation is sequential phosphorylation, where 
the sites are phosphorylated in a particular order and dephosphorylated in the 
reverse order - this is the case in the SH model. There is also a cyclic 
variant where dephosphorylation takes place in the same order as 
phosphorylation. It is also possible to consider phosphorylation in a random 
order or mixtures of the mechanisms introduced. An extensive discussion of the 
possibilities and examples of biological systems where they occur can be found 
in \cite{salazar09}. For many of these systems an analogue of Theorem 2 holds 
and can be proved in a similar way. For the only properties required for the 
proof are as follows:

\begin{itemize}
\item{each complex consists of one species}
\item{there is only one linkage class}
\item{the network is weakly reversible}
\end{itemize}
Thus the analogues of the SH model with cyclic or random phosphorylation
both have the property that there is a unique stationary solution in each 
stoichiometric compatibility class and that any other solution converges
to a stationary solution. 

There is another dynamical system related to the modelling of T cell activation
which has deficiency zero and can thus be shown to have the property that any
solution converges to a stationary solution and that there is only one 
stationary solution in any stoichiometric compatibility class. This is 
the kinetic proofreading model of McKeithan \cite{mckeithan95} for antigen 
recognition by the T cell receptor and its dynamics was analysed 
mathematically by Sontag \cite{sontag01}. In that paper the analogue of 
Theorem 2 is proved for McKeithan's model. The key observation is that the 
deficiency of the network is zero so that Theorem 1 applies. From there it
is possible to obtain the analogue of Theorem 2 for that system in a way very
similar to what has just been done for the SH model.

In the SH model each phosphorylation or dephosphorylation is modelled as
a single reaction and the details of the interaction with the enzyme which
catalyses the process are not included. Suppose that instead the enzyme is
incorporated in the standard Michaelis-Menten way \cite{murray}. This means 
that the reactions describing the formation of a complex of the substrate with 
the enzyme and the dissociation of the complex to give either enzyme and 
substrate or enzyme and product are included, using mass-action kinetics. This 
is what is 
referred to as Michaelis-Menten via mass action (MMvma) in \cite{grimbs11} and 
is different from using an effective Michaelis-Menten kinetics for a single 
reaction. The analogue of one of the parts of the decoupled SH model with 
the simple mass-action kinetics replaced by MMvma kinetics is similar to what 
is called a multiple futile cycle in \cite{wang08}. In that case there is only 
one kinase which catalyzes all phosphorylations and one phosphatase which 
catalyzes all dephosphorylations. This is slightly different from the 
situation in \cite{grimbs11}, where there is a different enzyme for each 
reaction. In \cite{wang08} upper and lower bounds for the number of stationary 
solutions of a system of this type are obtained. It follows from these that 
while in the case $N=1$  there is only one stationary solution there are at 
least three stationary solutions for $N=2$ and at least thirteen for $N=13$ 
for suitable choices of the parameters of the system. This corresponds to the 
case where the total concentrations of the enzymes are small compared to the 
total concentration of the substrates. The number of stationary solutions is 
never greater than $2N-1$, whatever the parameters. If the total 
concentrations of the enzymes are sufficiently large compared to the total 
concentrations of the substrates then there is at most one stationary solution. 

The equations arising in the decoupled system can be analysed in the same 
way as the full system and the analogue of Theorem 2 holds in that case.
For the decoupled system more can be done and the stationary solutions
can be calculated explicitly, as was shown in \cite{salazar03}. They are 
obtained by setting the sum of certain pairs of terms to zero. In the 
terminology of CRNT these stationary solutions are detailed balanced. Whether 
this gives the most general stationary solutions of the decoupled system is 
not discussed in \cite{salazar03} but it follows from the analogue of Theorem
2 that they are. To be concrete the system describing concentrations in the 
cytosol will be considered. To obtain the class of solutions found in 
\cite{salazar03} it is assumed that the first two terms on the second line of 
the evolution equation for $a_n$ cancel. This also gives a similar 
cancellation in the evolution equation for $i_n$. The condition for this is 
that $\frac{i_n}{a_n}=L_n$ where $L_n=\frac{l_n^-}{l_n^+}$. Next it is assumed 
that the second and third terms in the evolution equations for $a_n$ cancel 
for $0\le n\le N-1$, giving
\begin{equation}\label{balance}
\frac{a_{n+1}}{a_n}=\frac{k_n}{c_n}.
\end{equation}
Then the first and fourth terms cancel except in the case $n=0$. In fact,
if the equations involving $L_n$ are satisfied and the equations 
(\ref{balance}) are satisfied for all $n\le N-1$ then the conditions for a 
stationary solution of the part of the decoupled system describing 
concentrations in the cytosol is satisfied. Note that these detailed balance
conditions cannot be satisfied by a stationary solution of the full system
with non-zero coefficients $d$ and $f$. They may, however, be approximately 
satisfied when $d$ and $f$ are small. For these stationary solutions of
the decoupled system the fraction of the NFAT in the cytosol which is in the 
active state can be computed. It is given by
\begin{equation}\label{response}
\phi=\frac{\sum_{n=0}^N a_n}{\sum_{n=0}^N (a_n+i_n)}
=\frac{1+\sum_{n=1}^N\left(\prod_{j=0}^{n-1}\frac{k_j}{c_j}\right)}
{1+L_0+\sum_{n=1}^N(1+L_n)\prod_{j=0}^{n-1}\frac{k_j}{c_j}}.
\end{equation}  
In order to have a better understanding of the system it is useful to 
consider a special case with a reduced number of parameters. This is 
obtained in the following way. The coefficients $k_n$, $k_n'$, $c_n$ and 
$c_n'$ are taken to be independent of $n$ and denoted by $k$, $k'$, $c$
and $c'$ respectively. It is also assumed that $L_n=L_0\lambda^n$. In this
case the expression for $\phi$ becomes
\begin{equation}\label{response2}
\phi=\frac{\sum_{n=0}^N\left(\frac{k}{c}\right)^n}
{\sum_{n=0}^N(1+L_0\lambda^n)(\frac{k}{c})^n}
=\left[
1+L_0\frac{\left(\frac{\lambda k}{c}\right)^{N+1}-1}
{\frac{\lambda k}{c}-1}
\frac{\frac{k}{c}-1}{\left(\frac{k}{c}\right)^{N+1}-1}
\right]^{-1}.
\end{equation} 

One of the results of \cite{salazar03} is that for large $N$ the function 
$\phi$ resembles a Hill function $\phi_H(c)=\frac{c^N}{A+c^N}$ for a 
constant $A$ and an exponent $N$. In what sense does this resemblance hold? 
If $N$ is allowed 
to tend to infinity for fixed $A$ then the Hill function tends pointwise
almost everywhere to a translated Heaviside function which is zero for $c<1$
and one for $c>1$. From the point of view of the applications this gives rise
to a switch behaviour for large $N$. For values of the control parameter $c$
smaller than a threshold almost all the NFAT in the cytosol is in the inactive
form while for values larger than the threshold almost all the NFAT is in
the active form. The limiting behaviour of the function (\ref{response2}) as
$N$ tends to infinity depends on the assumptions made about the other 
parameters present. If the other parameters are fixed then what is obtained 
in the limit does contain a threshold but is not a switch. The amount of
activated NFAT is very small below the threshold but increases gradually 
above the threshold. This should be compared with the discussion in 
\cite{gunawardena05}.  Consider first the effect of varying $c$ while keeping 
the other parameters fixed. $\phi\to (1+L_0)^{-1}$ as $c\to\infty$ and 
$\phi\to (1+L_0\lambda^N)^{-1}$ as $c\to 0$. The quantity $\lambda$ is assumed 
to be greater than one in \cite{salazar03} and so if $N$ is large the value of 
$\phi$ at zero is close to zero. Consider next what happens if $N$ tends to 
infinity for fixed values of the other parameters. In the region where 
$c>\lambda k$ the inequality $\frac{k}{c}<1$ holds. Then the limit of $\phi$ 
as $N\to\infty$ is given by
\begin{equation}
\phi_\infty=\left[1+L_0\frac{\frac{k}{c}-1}{\frac{\lambda k}{c}-1}\right]^{-1}
=\frac{c-\lambda k}{c-\lambda k+L_0(c-k)}.
\end{equation}
When $\lambda k$ tends to $c$ the function $\phi_\infty$ tends to zero. Let 
$\tilde c=c-\lambda k$. Then
\begin{equation}
\phi_\infty=\frac{\tilde c}{(1+L_0)\tilde c+L_0k(\lambda-1)}
\end{equation}
for $c>\lambda k$. In the region where $c<\lambda k$ we get $\phi_\infty=0$.
Thus $\phi_\infty$ is a truncated translated Hill function with exponent one as 
mentioned in \cite{gunawardena05}.  There is also another interesting way of 
passing to the limit $N\to\infty$ which is more closely related to what is 
done in \cite{salazar03}. To see this it is convenient to introduce the 
variable $\mu=L_0^{\frac{2}{N}}\lambda$. Inverting this gives
$L_0=\left(\frac{\mu}{\lambda}\right)^{\frac{N}{2}}$. Substituting this into the 
expression for $\phi$ gives
\begin{equation}
\phi=\left[
1+\left(\frac{\lambda}{\mu}\right)^{\frac{N}{2}}
\frac{\left(\frac{k}{c}\right)^{N+1}
-\left(\frac{\mu}{\lambda}\right)^{N+1}}
{\frac{k}{c}-\frac{\mu}{\lambda}}
\frac{\frac{k}{c}-1}{\left(\frac{k}{c}\right)^{N+1}
-1}
\right]^{-1}.
\end{equation}
Denote the limit of this function as $\hat\phi_\infty$. Assume that 
$\frac{\mu}{\lambda}$ is a fixed number which is less than one. When 
$c>\sqrt{\frac{\lambda}{\mu}}k$ the function $\hat\phi_\infty$  is equal to 
one while when $c<\sqrt{\frac{\lambda}{\mu}}k$ it is equal to zero. Thus 
$\hat\phi_\infty$ is a translated Heaviside function and this limit behaves in 
a similar way to the limit of a Hill function when the exponent is allowed to 
tend to infinity while all other parameters are fixed. It is interesting to 
note that the threshold in $\hat\phi_\infty$ occurs in a different place from 
the threshold in $\phi_\infty$. In the example plotted in Fig. 2(b) of 
\cite{salazar03} the parameters are chosen as $\lambda=10$ and $\mu=1$. Then 
the threshold for $\hat\phi_\infty$ is at about $3.3$. At the threshold value 
of $c$ the function $\phi$ has exactly the value one half. One key property
implemented by this choice of parameters is that the reaction constants are 
such that there is a strong tendency for weakly phosphorylated NFAT to change 
from the inactive to the active conformation and for highly phosphorylated 
NFAT to change from the active to the inactive one. 

The validity of the above explicit formulae is restricted to the decoupled 
system but they do provide some information about the full system where $d$ 
and $f$ are non-zero.
When all other parameters are fixed there is a unique stationary solution
in a given reaction simplex for each choice of non-negative values of $d$ and
$f$. It can easily be shown, using the compactness of the reaction simplex,
that the stationary solution depends continuously on the parameters. Since
its position is known explicitly when $d=f=0$ its position is known 
approximately when these two parameters are small.

For any solution of the SH system let $\phi$ be the proportion of NFAT in the 
cytosol which is in the active state and $\psi$ the fraction of NFAT in the 
nucleus which is in the inactive state. Let $Z$ be the fraction of NFAT which 
is in the nucleus. Let $V_1$ and $V_2$ be the volume of cytosol and nucleus.
Then by the conservation of the total amount of NFAT the relation
\begin{equation}
fZ\psi V_2=d(1-Z)\phi V_1
\end{equation}
holds for any stationary solution and so
\begin{equation}
Z=\frac{d\phi V_1}{d\phi V_1+f\psi V_2}.
\end{equation}
If $f$ and $d$ small then we can obtain approximate expressions for $\phi$
and $\psi$ as functions of the reaction constants. Thus a corresponding
approximation is obtained for $Z$.

\section{Modelling the calcium influx}\label{calcium}

The model presented in the last section gives a description of how the
proportion $Z$ of active NFAT in the nucleus depends on the parameters
describing the state of the cell. An idealized
experiment would then consist in modifying the rate constants $c$ and $C$ 
by stimulating the cell and measuring the resulting change in $Z$. In 
real experiments things are more complicated. Consider now the experiments
carried out in \cite{podtschaske07}. There, among other things, the following 
type of experiment was done. A population of T cells was treated with 
different concentrations of ionomycin and the production of IL-2 by these
cells was measured. To get production of IL-2 the second signal is also
required and it is produced artificially by stimulating the cells with PMA
(phorbol 12-myristate 13-acetate). The result for a given level of 
stimulation is a curve describing the number of cells producing different 
amounts of IL-2. If all the cells were identical this curve would reduce to 
a Dirac $\delta$ but in reality this is smeared out to a smooth curve by 
the natural variability of the cells. When the stimulus is varied the 
following phenomenon is observed. The curve in general consists of two 
peaks separated by a region of production rates where there are very few 
cells. As the stimulus is varied the height of one of the peaks grows 
while that of the other shrinks. The value at which the maximum of a peak
is attained is essentially unchanged. The interpretation of this result is
that a given cell either produces almost no IL-2 or produces IL-2 at a close
to maximal rate. It exhibits a switch behaviour. This corresponds to the 
switch-like dependence of $\phi$ on $c$ found in the model.   

In the analysis of these results it is supposed that the number of cells where 
IL-2 production is switched on closely reflects the number of cells where 
$Z$ is almost one. On the other hand the idea that different levels of 
stimulation can be modelled by taking different constant values of $c$ and 
$C$ may be oversimplified and this issue will now be looked at in more detail.
The idealized situation would be to set a constant level of calcineurin 
activity by setting a constant calcium concentration. In fact it is observed
experimentally that when a calcium influx is caused by stimulating the TCR
and CD28, or by treating the cell with ionomycin and PMA, the concentration
of calcium often displays oscillations. It is also possible that these 
oscillations, rather than being an unimportant side effect, encode information.
A rise in calcium concentration does not only affect the NFAT signalling
pathway but also other characteristics of the cell. This raises the question
of how one signal can control several outputs. One possibility is that the 
information is encoded in the time dependence of the calcium concentrations
and that the oscillations have an important role to play in this. Some 
experimental results on calcium oscillations in liver cells and T cells are
described in \cite{somogyi91} and \cite{dolmetsch94} respectively. For an
extensive review of modelling of calcium dynamics and signalling see 
\cite{falcke04}. 

The constants $c$ and $C$ represent the concentration of active calcineurin
and so in modelling the effects of calcium influx they should be replaced
by functions of time. How does the concentration of active calcineurin
depend on the concentration of calcium? In \cite{salazar03} this is modelled
by an equation of the form
\begin{equation}\label{calcineurin}
\frac{dz}{dt}=a(z_0-z)y-bz
\end{equation}
where $z$ is the concentration of active calcineurin, $y$ is the 
concentration of calcium in the cytosol and $a$ and $b$ are constants. It 
remains to model the dependence of the calcium concentration on time, possibly 
including oscillations. In \cite{salazar03} this is done by means of a system 
of two ordinary differential equations. These equations contain the 
concentration of IP${}_3$. With a view to modelling stimulation by ionomycin 
it will be assumed that the concentration of IP${}_3$ is constant. Then the 
model for the calcium concentration is schematically of the following form:
\begin{eqnarray}
&&\dot x=\rho [-\alpha (x-y)+\beta y-\lambda f(y)(x-y)],\label{ssmodel1}\\
&&\dot y=\alpha (x-y)-\beta y+\lambda f(y)(x-y)+\gamma-\delta y.\label{ssmodel2}
\end{eqnarray}
Here $x$ is the concentration of calcium in the lumen of the endoplasmic 
reticulum. The quantities $\alpha$, $\beta$, $\gamma$, $\delta$, $\lambda$ 
and $\rho$ are positive constants and $f$ is a positive function describing 
the response 
of the IP${}_3$ receptor to the calcium concentration. In \cite{salazar03} it 
is chosen to be a Hill function with exponent two. This system is essentially 
a special case of one introduced by Somogyi and Stucki \cite{somogyi91}. The
only reason it is not included is that the model in \cite{somogyi91} has 
$\rho=1$. Thus it implicitly assumes that the cytosol and the lumen of the 
endoplasmic reticulum have the same volume. In \cite{somogyi91} 
another variant was also considered where the fact that $y$ is much smaller 
than $x$ is used to replace $x-y$ by $x$ wherever it occurs in the above 
system. The above system will be called the modified Somogyi-Stucki model 
while the other variant will be called the unmodified Somogyi-Stucki model. 
Note that in the unmodified system a multiplicative constant like $\rho$ 
can be removed by rescaling the variable $x$ and some of the parameters. It is 
remarked in \cite{somogyi91} that taking $f(y)=y^2$ and $\alpha=0$ in the 
unmodified model gives a system which is equivalent to the well-known 
Brusselator \cite{prigogine68}.

Now some remarks will be made on the dynamics of solutions of the system 
(\ref{ssmodel1})-(\ref{ssmodel2}). Note first that the only terms on the 
right hand side of the evolution equation for a given unknown which are 
negative contain a factor of that unknown. Thus it can be shown as in the 
proof of Lemma 1 that a solution which starts positive remains positive. 
Taking a linear combination of the two evolution equations shows that any 
stationary solution $(x_*,y_*)$ satisfies $y_*=\frac{\gamma}{\delta}$. 
Substituting this back in the equation $\dot x=0$ gives 
\begin{equation}
x_*=\frac{\gamma [\alpha+\beta+\lambda f(y_*)]}{\delta [\alpha+\lambda f(y_*)]}.
\end{equation}
Thus, in particular the system has exactly one stationary solution for any 
choice of the parameters and the function $f$. Linearizing the system
about the stationary solution gives
\begin{eqnarray}\label{lin}
&&\frac{d\tilde x}{dt}=\rho\{[-\alpha-\lambda f(y_*)]\tilde x
+[\alpha+\beta-\lambda f'(y_*)(x_*-y_*)+\lambda f(y_*)]\tilde y\},\\
&&\frac{d\tilde y}{dt}=[\alpha+\lambda f(y_*)]\tilde x
+[-\alpha-\beta+\lambda f'(y_*)(x_*-y_*)-\lambda f(y_*)-\delta]\tilde y.
\end{eqnarray}
The determinant of the linearization at $(x_*,y_*)$ is 
$\delta\rho(\alpha+\lambda f(y_*))>0$. Thus 
the linearized stability of the stationary point is determined by the sign 
of the trace of the linearization which is given by
\begin{equation}\label{trace}
-(1+\rho)\alpha-\beta-\delta-(1+\rho)\lambda f(y_*)+\lambda f'(y_*)(x_*-y_*).
\end{equation} 
The stationary solution is a hyperbolic source if and only if
\begin{equation}
\rho<(\alpha+\lambda f(y_*))^{-1}[-\alpha-\beta-\delta-\lambda f(y_*)
+\lambda f'(y_*)(x_*-y_*)].
\end{equation}
Since $x_*$ and $y_*$ do not depend on $\rho$ this shows that the region of 
instability is non-empty and is bounded by a smooth hypersurface in parameter 
space.

In the model of $\cite{salazar03}$ the nonlinearity is given by 
$f(y)=\frac{y^2}{A+y^2}$. It follows that $f'(y)=\frac{2Ay}{(A+y^2)^2}$ and
so  
\begin{eqnarray}\label{largelambda}
&&-(1+\rho)f(y_*)+f'(y_*)(x_*-y_*)\nonumber\\
&&=-(1+\rho)\frac{y_*^2}{A+y_*^2}
+\frac{2Ay_*}{(A+y_*^2)}\frac{\beta\gamma}{\delta[\alpha(A+y_*^2)
+\lambda y_*^2)]}
\nonumber\\
&&=\frac
{\{-(\alpha+\lambda)(1+\rho)y_*^2+A[-(1+\rho)\alpha+2\beta]\}y_*^2}
{(A+y_*^2)[\alpha (A+y_*^2)+\lambda y_*^2]}.
\end{eqnarray}
This means in particular that in this case a necessary condition for the 
positivity of the trace is 
\begin{equation}\label{largelambdaineq}
\left(\frac{\gamma}{\delta}\right)^2<\frac{A[2\beta-(1+\rho)\alpha]}
{(\alpha+\lambda)(1+\rho)}.
\end{equation}
 
Now it will be proved that every solution of the Somogyi-Stucki model is
bounded. Note first that
\begin{equation}
\frac{d}{dt}(x+\rho y)=\rho(\gamma-\delta y).
\end{equation}
Let $L_C$ be the part of the line $x+\rho y=C$ for which $y\ge\gamma/\delta$. 
No solution can cross it in the direction of increasing $y$. Let $L'$ be the 
line $y=\frac{\gamma}{\alpha+\beta+\delta+\lambda}$. It cannot be crossed 
in the direction of decreasing $y$. Let $H$ be the zero set of the right
hand side of (\ref{ssmodel1}). Its equation can be written as
\begin{equation}
x=y\left[1+\frac{\beta}{\alpha+\lambda f(y)}\right].
\end{equation}
This curve starts at the origin and as $x$ tends to infinity along it $y$ 
tends to infinity. Let
\begin{equation}
M=\frac{\gamma}{\delta}\left[1+\frac{\beta}{\alpha+\lambda 
f(\frac{\gamma}{\delta})}\right].
\end{equation}
Then the curve crosses the line $y=\gamma/\delta$ when $x=M$.
Define a region $R_C$ to be that bounded by the $y$-axis, the line
$L_C$, the vertical line joining the endpoint of $L_C$ with $L'$
and the part of $L'$ between its intersection with that vertical line
and the $y$-axis. For $C\ge M+\frac{\rho\gamma}{\delta}$ this region is 
invariant under the flow. Hence any solution which starts in $R_C$ remains 
there and is bounded. Since the union of the allowed $R_C$ cover the region 
$y\ge\frac{\gamma}{\alpha+\beta+\delta+\lambda}$ it follows that any 
solution which tended to infinity would have to lie in the strip
$y<\frac{\gamma}{\alpha+\beta+\delta+\lambda}$. Hence for a solution which is 
unbounded $y>M$ implies $\dot y<0$, a contradiction, and 
all solutions are bounded. It follows from Poincar\'e-Bendixson theory that 
unless the trace of the linearization at the stationary solution vanishes
each solution either converges to a stationary solution or to a periodic 
solution. The condition on the trace is required so as to rule out possible
orbits homoclinic to the stationary solution.

For certain ranges of the parameters it can be shown that there are no 
periodic solutions. To see this note first that any periodic solution must
lie in the region $y>\frac{\gamma}{\alpha+\beta+\delta+\lambda}$. For it can 
neither lie entirely in the complement of that region, due to the monotonicity 
of $y$ there, nor leave that region. It must lie in some $R_C$ and in fact it
must do so for some $C\le M+\frac{\rho\gamma}{\delta}$. For the boundary of 
$R_C$ with any larger value of $C$ can only be crossed in one direction. 
In particular for this solution $x$ can never exceed 
$M$. Consider the divergence of the vector 
field defining the dynamical system. If the divergence is negative on a region 
containing the solution being considered then a contradiction is obtained. 
This divergence is given by replacing $x_*$ and $y_*$ in (\ref{trace}) by $x$
and $y$ respectively and can be written in the form
\begin{equation}
-[(\rho+1)\alpha+\beta+\delta+(\rho+1)\lambda f(y)+\lambda f'(y)y]+\lambda
f'(y)x.
\end{equation}
The last term is the only positive one and it can be bounded above by 
$\frac{3\sqrt{3}\lambda M}{8A^{\frac32}}$ when $x\le C$. Hence if
\begin{equation}\label{noperiodic}
\frac{3\sqrt{3}\lambda M}{8A^{\frac32}}\le (\rho+1)\alpha+\beta+\delta
\end{equation}
the divergence is negative on $R_C$. It follows that a sufficient condition
for the absence of periodic solutions is that
\begin{equation}\label{noperiodic2}
\frac{3\sqrt{3}\lambda\gamma}{8A^{\frac32}\delta} \left(1
+\frac{\beta}{\alpha+\lambda f(\frac{\gamma}{\delta})}\right)
\le (\rho+1)\alpha+\beta+\delta.
\end{equation}
When there are no periodic solutions it can be concluded that every solution 
converges to the stationary solution. 

In \cite{salazar03} numerical simulations were done for the above model with
certain values of the parameters. These are:
\begin{equation}
\alpha=0.001,\ \gamma=0.38,\ \delta=1,\ \lambda=0.16,\ 
\rho=10,\ A=0.25.
\end{equation}
In two different simulations $\beta$ was chosen to be $0.1$ and $1$. In both 
these cases the quantity (\ref{trace}) is negative and so the stationary
solution is a hyperbolic sink. For $\beta=0.1$ the inequality 
(\ref{largelambdaineq}) is violated and so this necessary condition suffices to
verify that the stationary solution is a sink in that case. For $\beta$ 
sufficiently large with the other parameters as above the trace is positive. 
The value of $\beta$ for which the trace vanishes is about $8.2$. For 
$\beta=0.1$ the inequality (\ref{noperiodic2}) is satisfied and the criterion 
for the absence of periodic solutions applies. For $\beta=1$ it does not. 

Suppose that for certain values of the parameters in 
(\ref{ssmodel1})-(\ref{ssmodel2}) and 
certain initial data the solution converges exponentially to the stationary 
solution $(x_*,y_*)$ as $t\to\infty$. In other words, assume there are
positive constants $\eta$ and $R$ such that
\begin{equation}
|x(t)-x_*|+|y(t)-y_*|\le Re^{-\eta t}.
\end{equation} 
The equation (\ref{calcineurin}) can be rewritten as
\begin{equation}
\frac{dz}{dt}=-(b+ay)z+az_0y.
\end{equation}
Define
\begin{equation}
z_*=\frac{az_0y_*}{b+ay_*}.
\end{equation}
Then it follows from the variation of constants formula that
\begin{equation}
z(t)=z(t_0)e^{-b(t-t_0)}e^{-a\int_{t_0}^ty(s)ds}
+\int_{t_0}^t e^{-b(t-s)}e^{-a\int_s^t y(u)du}az_0y(s)ds
\end{equation} 
and hence that
$z(t)=z_*+\ldots$ where the remainder decays exponentially as $t\to\infty$.

What can be said about the asymptotic behaviour of solutions of the system 
obtained when the SH model is generalized by replacing the constants $c$ and
$C$ by functions of $t$ which converge to constants $c_*$ and $C_*$ 
respectively as $t\to\infty$? Denote by $X(t)$ a solution of the system of ODE
obtained from the SH model by allowing the coefficients to be time dependent.
Denote this system by $\dot X=F(X)$. Suppose that these coefficients converge 
to positive limits as $t\to\infty$ and that their time derivatives tend to 
zero. If $t_n$ is a sequence tending
to infinity as $n\to\infty$ consider the translates $X_n(t)=X(t+t_n)$. Because 
of the compactness of the reaction simplex these are uniformly bounded. Using 
this in the evolution equations shows that the time derivatives $X'_n(t)$ are 
also uniformly bounded. Differentiating the equation shows that the sequence
$X''_n(t)$ is uniformly bounded. By the Arzela-Ascoli theorem there is a 
subsequence $X_{n_r}$ which converges together with its first derivatives 
uniformly on compact subsets to a limit $X_\infty(t)$. The functions $X_n$
satisfy
\begin{equation}
\dot X_n(t)=F(X_n(t))=F(X(t+t_n))=F_\infty(X_n(t))+\ldots.
\end{equation} 
Here $F_\infty$ is obtained by replacing the coefficients in $F$ by their 
limiting values and the remainder term converges to zero as $n\to\infty$.
Hence
\begin{equation}
\dot X_\infty(t)=F_\infty(X_\infty(t)) 
\end{equation}
and $X_\infty(t)$ solves the SH system. All solutions of the SH system tend to
the same limiting value $X_*$. Thus by passing to a subsequence once more it 
can be seen that each sequence $\{t_n\}$ has a subsequence along which $X(t)$
converges to $X_*$. Thus in fact $X(t)\to X_*$.

If a solution of (\ref{ssmodel1})-(\ref{ssmodel2}) does not converge to a 
stationary solution then it must converge to a periodic solution. Moreover, 
there are solutions where this happens whenever the parameters are such that 
the stationary point $(x_*,y_*)$ is unstable. In this case it is more 
difficult to understand the behaviour of $z$ and of the modified version of 
the SH model with variable $c$ and $C$. 

\section{Conclusions and outlook}\label{conclusion}

In this paper the properties of some mathematical models of parts of the 
NFAT signalling pathway have been investigated. One of the main results 
concerns a model describing the phosphorylation states of the transcription
factor NFAT and the transport of these between the cytosol and the nucleus.
Using chemical reaction network theory it was shown that for any value of 
the parameters in this system every solution converges to a stationary 
solution as $t\to\infty$ and that this stationary solution is uniquely 
determined by the total amount of NFAT in the cell. Moreover it was shown 
that in the case where transport between the two compartments is slow compared 
to the reactions within each compartment the stationary solution can be 
approximated by explicit expressions derived in \cite{salazar03}. It was 
exhibited in which way this model can behave as a switch.

The parameters in the model describing the phosphorylation states include
some which reflect the concentration of active calcineurin in the cytosol.
It is a priori not clear that it is sufficient to assume that this 
concentration is constant and for this reason a model allowing a
time-dependent calcineurin concentration was examined. The concentration
of active calcineurin depends on the calcium concentration in the cytosol
and this concentration is affected when a T cell is activated. The core 
model studied is a two-dimensional dynamical system closely related to
a model for calcium dynamics introduced in \cite{somogyi91}. Solutions 
of this model become periodic at late times for some values of the parameters
and converge to a constant at late times for others. Criteria were obtained 
for the parameter values leading to these two different outcomes. It was
shown that when the solution converges to a constant at late times this leads
to a calcineurin concentration which also becomes constant and that
the resultant densities of the phosphorylation states of NFAT in these cases
are as in the case of constant calcineurin concentrations.

In the course of this work the following additional questions arose. The
criteria obtained for when exactly the system (\ref{ssmodel1})-(\ref{ssmodel2})
does or does not have periodic solutions are far from exhaustive. In 
particular, one question left open is if there are parameter values for 
which the stationary solution is stable but there nevertheless exist
periodic solutions. It was shown how the SH model is affected by an 
asymptotically constant input. It would be interesting to know how it is 
affected by an asymptotically periodic input. More generally it may be 
asked what can be said about a possible generalization of CRNT where 
the reaction constants are replaced by periodic functions.

It was remarked that the results obtained for a model like the SH model, where 
mass action kinetics are used, may change essentially if another type of
kinetics, such as Michaelis-Menten via mass action, is used. What happens 
when this is replaced by effective Michaelis-Menten kinetics? Here we are
talking about different ways of modelling the same system of chemical 
reactions and it should be possible to relate them in a mathematically
rigorous way.  
 
Ultimately, in understanding T cell activation, the interaction of the
NFAT signalling pathway with the signalling pathways leading to other 
important transcription factors should be understood. Is it enough, when 
examining the coupled network, to confine attention to stationary solutions?
Or so more complicated dynamical phenomena play a role? That they may do so 
is suggested by the results of \cite{dolmetsch97} where it is shown that 
for instance NFAT and NF$\kappa$B react to time-dependent calcium signals in 
different ways. A first step towards doing this could be to examine the
influence of the dynamical behaviour within each of the individual 
pathways. One dynamical system describing NF$\kappa$B signalling is given
in \cite{fisher06}. The signal pathway leading from the T cell receptor to
AP-1 passes through a MAP-kinase cascade. Mathematical modelling of this 
pathway revealing switch-like behaviour has been carried out in 
\cite{altanbonnet}. An insightful review of the possible contributions
of theory and computation to the understanding of signalling pathways, with
particular attention to T cells, has recently been given by Chakraborty and
Das \cite{chakraborty}. 

\vskip 10pt 
\noindent
{\it Acknowledgements} I am grateful to Ria Baumgrass and Bernold Fiedler for
helpful discussions.

\appendix
\section{Remarks on the model of Fisher et. al.}

The purpose of this appendix is to discuss the relation of the SH model 
discussed in the main part of the text to a model introduced by Fisher et. al.
in \cite{fisher06}. The building blocks in the latter model are NFAT and 
calcineurin. The NFAT may be unphosphorylated or fully phosphorylated and
intermediate states are not included. The calcineurin may be active or 
inactive and the active form may bind to the different forms of NFAT.
Concentrations of all these substances in the cytosol and the nucleus are
considered and all of them may be transported between the two compartments.
There are twelve concentrations in all. They satisfy a system of equations
which, in neutral notation, can be written as:
\begin{eqnarray}
&&\frac{dx_1}{dt}=k_1x_5-k_2x_1+k_{17}\frac{v_c}{v_n}x_2-k_{18}x_1
+k_{15}x_9-k_{16}x_1x_3\nonumber\\
&&\frac{dx_2}{dt}=k_1x_6-k_2x_2+k_{18}\frac{v_n}{v_c}x_1-k_{17}x_2
+k_{15}x_{10}-k_{16}x_2x_4\nonumber\\
&&\frac{dx_3}{dt}=-k_{11}x_5x_3+k_{12}x_7+k_5\frac{v_c}{v_n}x_4
-k_6x_3+k_{15}x_9-k_{16}x_3x_1+k_{19}Ix_{11}-k_{20}x_3\nonumber\\
&&\frac{dx_4}{dt}=-k_{11}x_6x_4+k_{12}x_8-k_5x_4+k_6\frac{v_n}{v_c}x_3
+k_{15}x_{10}-k_{16}x_4x_2+k_{19}Ix_{12}-k_{20}x_4\nonumber\\
&&\frac{dx_5}{dt}=-k_1x_5+k_2x_1-k_4x_5+k_3\frac{v_c}{v_n}x_6
-k_{11}x_5x_3+k_{12}x_7\nonumber\\
&&\frac{dx_6}{dt}=-k_1x_6+k_2x_2-k_3x_6+k_4\frac{v_n}{v_c}x_5
-k_{11}x_6x_4+k_{12}x_8\nonumber\\
&&\frac{dx_7}{dt}=k_{11}x_5x_3-k_{12}x_7+k_7\frac{v_c}{v_n}x_8
-k_8x_7-k_{13}x_7+k_{14}x_9\nonumber\\
&&\frac{dx_8}{dt}=k_{11}x_6x_4-k_{12}x_8+k_8\frac{v_n}{v_c}x_7
-k_7x_8-k_{13}x_8+k_{14}x_{10}\nonumber\\
&&\frac{dx_9}{dt}=k_{16}x_1x_3-k_{15}x_9+k_{13}x_7-k_{14}x_9
+k_9\frac{v_c}{v_n}x_{10}-k_{10}x_9\nonumber\\
&&\frac{dx_{10}}{dt}=k_{16}x_2x_4-k_{15}x_{10}+k_{13}x_8-k_{14}x_{10}
+k_{10}\frac{v_n}{v_c}x_9-k_9x_{10}\nonumber\\
&&\frac{dx_{11}}{dt}=k_5\frac{v_c}{v_n}x_{12}-k_6x_{11}-k_{19}Ix_{11}
+k_{20}x_3\nonumber\\
&&\frac{dx_{12}}{dt}=-k_5x_{12}+k_6\frac{v_n}{v_c}x_{11}-k_{19}Ix_{12}
+k_{20}x_4\nonumber
\end{eqnarray}
Here all quantities other than the unknowns $x_i$ are positive constants.
To see the relation of these equations to those of \cite{fisher06} it suffices
to note that they are written in the same order in both cases. There are
conserved quantities
\begin{equation}
J_1=v_n(x_1+x_5+x_7+x_9)+v_c(x_2+x_6+x_8+x_{10}).
\end{equation}
and 
\begin{equation}
J_2=v_n(x_3+x_7+x_9+x_{11})+v_c(x_4+x_8+x_{10}+x_{12}).
\end{equation}
$J_1$ is the total concentration of NFAT in all forms and $J_2$ the total
concentration of calcineurin.

These equations do not appear to incorporate any switch-like mechanism and 
so it is difficult to see how they could model the behaviour observed in
\cite{podtschaske07}. The network structure contains rather little 
information about the nature of the substances involved and so it is to
be expected that important information influencing the dynamics is encoded in
the numerical values of the reaction constants $k_i$. For instance, the 
fact that the central role of calcineurin in this context is that of a 
phosphatase is not visible from the equations. In \cite{fisher06} specific
experimentally motivated values are chosen for the $k_i$ and the character
of calcineurin as a phosphatase is encoded in the fact that $k_{13}$ is 
much greater than $k_{14}$. In view of this it is perhaps not surprising that
CRNT is not too helpful in this case. There are sixteen complexes and only one 
linkage class. Since the rank of $\bar N$ cannot be more than the number of 
species, which is twelve, the deficiency must be at least three. Thus this 
system is far from the low deficiency situation where CRNT is most powerful.
Nevertheless a couple of simple conclusions can be drawn.

Because of the conserved quantities the reaction simplex is compact and all
solutions exist globally in time. Suppose that a solution with $J_1>0$ and 
$J_2>0$ has an $\omega$-limit point on the boundary of the positive quadrant.
Then arguing as in Section 2 we can conclude that there is a solution
$c_*$ lying in the boundary and that at late times it has a constant number 
of non-zero components. Let $n_1$ be the number of non-zero components and
$n_0=n-n_1$. Let ${\cal S}_1$ and ${\cal S}_0$ be the corresponding subsets 
of ${\cal S}$. It is impossible that there is a reaction whose left hand side 
is a species belonging to ${\cal S}_1$ and whose right hand side includes a 
species belonging to ${\cal S}_0$. Inspecting the pairs of species where
there are reactions of this type in both directions shows that each of the 
following three sets are subsets of either ${\cal S}_0$ or  ${\cal S}_1$
\begin{equation}
{\cal S}'_1=\{x_1,x_2,x_5,x_6\}, {\cal S}'_2=\{x_3,x_4,x_{11},x_{12}\}, 
{\cal S}'_3=\{x_7,x_8,x_9,x_{10}\}.
\end{equation} 
Moreover if ${\cal S}'_3\subset {\cal S}_1$ then ${\cal S}_1={\cal S}$.
This possibility is ruled out by the fact that the solution lies on 
the boundary of the positive orthant. Inspecting the system once again and 
using the form of the nonlinear terms shows that if both ${\cal S}'_1$ and
${\cal S}'_2$ are subsets of ${\cal S}_1$ a contradiction is obtained.
Thus the only remaining possibilities are ${\cal S}_1={\cal S}'_1$ of
${\cal S}_1={\cal S}'_2$. However each of these implies the vanishing of one
of the conserved quantities. It can be concluded that a solution for 
which $Z_1$ and $Z_2$ are non-zero has no $\omega$-limit points on the 
boundary. 

It follows from the Brouwer fixed point theorerm as in Theorem I.8.2
of \cite{hale} that there exists a stationary solution in the reaction 
simplex. Since it has been shown that this solution does not lie on the 
boundary it must lie in the interior. Thus this system has at least one
positive stationary solution for any choice of the parameters and 
any biologically relevant initial data. On the other hand this argument 
gives no indication as to whether there might be multiple stationary 
solutions or more complicated dynamics such as periodic solutions.

\end{document}